\documentclass[prl,aps,twocolumn,showpacs,amssymb]{revtex4}

\usepackage{graphics}
\usepackage{float,epsfig}

\begin{document}

\title{Proton-neutron  multiplets in exotic $^{134}$Sb: testing the shell-model effective interaction}
\author{L. Coraggio} 
\author{A. Covello} 
\author{A. Gargano} 
\author{N. Itaco}
\affiliation{Dipartimento di Scienze Fisiche, Universit\`a
di Napoli Federico II, \\ and Istituto Nazionale di Fisica Nucleare, \\
Complesso Universitario di Monte  S. Angelo, Via Cintia - I-80126 Napoli,
Italy}

\date{\today}

\begin{abstract}
The experimental level structure of $^{134}$Sb is compared with the results of a shell-model
calculation in which the two-body matrix elements of the effective interaction have been derived 
from the CD-Bonn nucleon-nucleon potential. The experimental data,
including the very low-lying first-excited $1^{-}$ state, are remarkably well reproduced by the theory. 
The results of this paper complement those
of our previous studies on $^{135}$Sb and $^{134}$Sn, showing that our effective interaction is well suited to
describe $^{132}$Sn neighbors beyond $N=82$. The various terms which contribute to 
the matrix elements of the proton-neutron effective interaction are examined and  
their relative importance is evidenced.   

\end{abstract}    
\pacs{21.60.Cs, 21.30.Fe, 27.60.+j}

\maketitle

The study of nuclei in the regions of shell closures off stability is currently the subject of great
interest, with 
particular attention focused on the neighbors of doubly magic $^{132}$Sn. A considerable effort has been  recently 
made to gain information on neutron-rich nuclei 
beyond the $N=82$ shell closure. The data which are  becoming available appear to be somewhat different from what 
one might expect by extrapolating the existing results for $N<82$ nuclei. For instance,
the first $2^{+}$ state in $^{134}$Sn lies  at 726 keV excitation energy, which makes it the lowest 
first-excited $2^{+}$ level observed
in a semi-magic even-even nucleus over the whole chart of nuclides. A significant drop in the energy of the lowest-lying 
$5/2^{+}$ state in $^{135}$Sb has been observed as compared to the values measured for the Sb isotopes with $N \leq 82$. 
In $^{134}$Sb, the $0^{-}$ ground state
and the first excited $1^{-}$ state are nearly degenerate, the latter lying at 13 keV.
This situation is similar to what occurs in  $^{210}$Bi, which is the counterpart of $^{134}$Sb in the lead region. 
In this case, the lowest multiplet, 
$\pi h_{9/2} \nu g_{9/2}$, shows a breakdown of the Nordheim strong rule \cite{nord} in that the $1^{-}$ state becomes
 the ground state with the $0^{-}$ state at about 50 keV excitation energy.

These new data might  be seen as  the onset of a modification in the shell structure,
which, starting at  $N=83-84$ , is expected to produce more evident 
effects for larger neutron excess. As an example, a possible explanation of the position 
of the $5/2^{+}$ state in $^{135}$Sb may reside in 
a downshift of the $d_{5/2}$ proton level relative to the $g_{7/2}$ one caused by a more diffuse nuclear 
surface produced by the two neutrons beyond the 82 shell closure. A shell-model calculation using
experimental single-particle (SP) energies with the above downshift set at 300 keV
and a two-body effective interaction derived from a modern nucleon-nucleon ($NN$) potential
leads indeed to a good description of  $^{135}$Sb \cite{sher02,sher05a}. However, 
while this Hamiltonian also provides a satisfactory agreement with experiment for $^{134}$Sn \cite{sher02}
this is not the case for $^{134}$Sb \cite{sher05}.  Other  shell-model calculations 
have been performed
for these three nuclei \cite{chou92, sark01,korg02}, making 
use of effective interactions  obtained through various different
modifications of an interaction \cite{warb91}  which was  originally constructed for the $^{208}$Pb region 
starting from  the Kuo-Herling matrix elements \cite{kuo71}. 
None of these calculations, however,  is able to account simultaneously for the
peculiar features of $^{134}$Sn, $^{134}$Sb, and $^{135}$Sb.
 
In this context, therefore, it is quite a relevant question whether there is a unique consistent Hamiltonian 
able to do this and, in particular, whether it may be derived from a realistic  free $NN$ potential. More precisely,
using SP energies taken from the experiment and a
two-body effective interaction derived from a modern  $NN$ potential, we would like to see how few-valence-particle 
nuclei just above $^{132}$Sn are described so as to ascertain  whether effects beyond the standard shell model are
really indispensable.

Very recently we have studied $^{135}$Sb \cite{cor05} and $^{134}$Sn \cite{cove} within the 
framework of the  shell model   
assuming $^{132}$Sn as a closed core and taking  as model space for the valence proton and neutrons 
the five  levels  $0g_{7/2}$, $1d_{5/2}$, $1d_{3/2}$, $2s_{1/2}$  
and $0h_{11/2}$ of the 50-82 shell and  
the six levels $0h_{9/2}$, $1f_{7/2}$, $1f_{5/2}$, $2p_{3/2}$,
$2p_{1/2}$, and  $0i_{13/2}$ of the 82-126 shell, respectively.
The two-body effective interaction has been derived by means of a $\hat Q$-box folded-diagrams method \cite{ei}
from the CD-Bonn $NN$ potential \cite{mac01}, the short-range repulsion of the latter being 
renormalized by use of the low-momentum potential $V_{\rm low-k}$ \cite{bog02}.
For protons, the Coulomb interaction has been added to $V_{\rm low-k}$.
As for the proton and neutron SP energies, they  have been taken from the experimental spectra of $^{133}$Sb and $^{133}$Sn,
respectively, with the exception of  the proton $s_{1/2}$ and the neutron $i_{13/2}$ levels which are still missing.
The values of $\epsilon_{s_{1/2}}$ and $\epsilon_{i_{13/2}}$  have been taken from Refs. \cite{and97} and \cite{cor02}, respectively, 
where it is discussed how they are determined.
All the  adopted values of the SP energies are reported in \cite{cor05}, 
to which we also refer for  a brief discussion of our derivation 
of the two-body effective interaction.     

Here we report  on  our study  of  $^{134}$Sb, which has been performed  using the same Hamiltonian as above.   
The calculations have been carried out by using the OXBASH shell-model code 
\cite{oxba}.

\begin{figure}[H]
\begin{center}
\includegraphics[scale=0.6,angle=0]{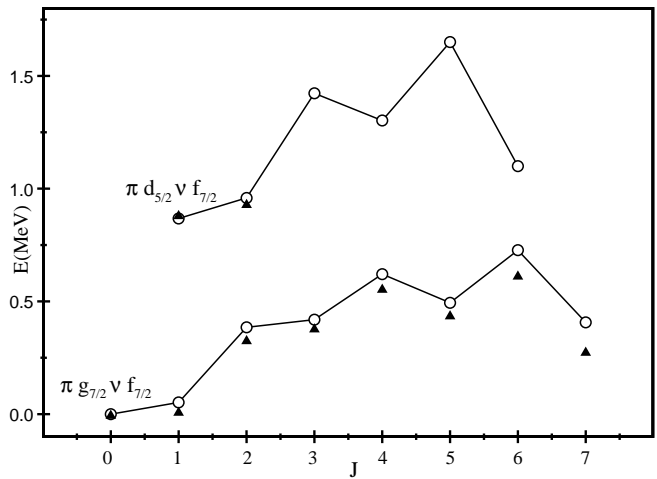}
\end{center}
\caption{Proton-neutron $\pi g_{7/2} \nu f_{7/2}$ and   $\pi d_{5/2}
  \nu f_{7/2}$ multiplets in $^{134}$Sb. The theoretical results are
  represented by open circles while the experimental data by solid
  triangles.}
\end{figure}

We now present our results starting  with  the binding energy of
the ground state. Our calculated value is $12.85 \pm0.05$ MeV, which compares very well 
with  the experimental one,  
$12.96 \pm 0.05$ MeV \cite{audi03}.  
Note that the error on the calculated value arises from the experimental errors on  
the proton and neutron separation energies of $^{133}$Sb and $^{133}$Sn \cite{foge99}.

The calculated energies of the $\pi g_{7/2} \nu f_{7/2}$ and   $\pi d_{5/2} \nu f_{7/2}$ 
multiplets in $^{134}$Sb are reported in Fig. 1, where they are compared with the  
experimental data \cite{sher05,nndc}. The first eight calculated states arise from
the $\pi g_{7/2} \nu f_{7/2}$ configuration and have  their experimental
counterpart in the eight lowest-lying experimental states. 
The wave functions of these states are characterized  by 
very little configuration mixing, the percentage of the leading component having a minimum value
of 88\% for the $J^{\pi}=2^{-}$ state 
while ranging from 94\% to 100\% for all other states.
As for the  $\pi d_{5/2} \nu f_{7/2}$ multiplet, we find that the
$1^{-}$, $2^{-}$,  $4^{-}$, and  $6^{-}$ members correspond to the
yrare states, while both the other two, with $J^{\pi}= 3^{-}$ and  $5^{-}$, to the third  
excited state.
In fact, the second  $3^{-}$ and  $5^{-}$ states, which are predicted at 1.42 and 1.46 MeV 
excitation energy, belong to the $\pi g_{7/2} \nu p_{3/2}$ configuration.  
As is shown in Fig. 1, only the  $1^{-}$ and  $2^{-}$
members of the $\pi d_{5/2} \nu f_{7/2}$ multiplet are known. Actually, a $(5^{-})$  state  
at 1.38 MeV  has been observed which we identify, however, with our second $5^{-}$ state. 
As regards the structure of the states belonging  to the $\pi d_{5/2} \nu f_{7/2}$ multiplet, we find that all 
members  receive significant 
contributions from configurations other than the dominant one, their percentage reaching even   
50\% in the case of the $2^-$ state.

We see that the agreement between theory and experiment is very good, the  discrepancies being in the order of a few tens
of keV for most of the states. The largest discrepancy occurs for the $7^{-}$ state, which lies at about 130 keV 
above its experimental counterpart. It is an important outcome of our calculation  that we
predict  almost the right spacing between the $0^{-}$ ground state and first excited 
$1^{-}$ state. In fact, the latter has been observed at 13 keV excitation energy, 
our value being 53  keV. 
It is worth mentioning that in the preliminary calculations of Ref. \cite{garga04} we 
overestimated the excitation energy of the $1^{-}$ state
by about 200 keV. Comments on this point can be found in Ref. \cite{cor05}.

In the recent paper mentioned above, Ref. \cite{sher05}, calculations have been performed for $^{134}$Sb using an effective 
interaction derived from the CD-Bonn-96  $NN$ potential by means of a
$G$-matrix folded-diagram method including diagrams up to third order \cite{brown05}. From comparison  
with experiment
for the eight states of the $\pi g_{7/2} \nu f_{7/2}$ multiplet, an average deviation of about
150 keV was found, and, in particular, the $1^{-}$ state turned  out to lie at 329 keV above the $0^{-}$ ground state.
As for the $\pi d_{5/2} \nu f_{7/2}$ multiplet, while the $2^{-}$ state is overestimated by about 120 keV the 
discrepancy for the $1^{-}$ state reaches about 380 keV. In the same paper, results are also presented which have been 
obtained with different two-body matrix elements, namely those of the KH208 \cite{warb91} and KH5082 interactions \cite{chou92}.
Calculations for $^{134}$Sb with the latter interaction were already performed in Ref. \cite{korg02}.
The authors of \cite{sher05} present the results of both calculations pointing out  
that they are in  better agreement with the experimental data than those obtained with the CD-Bonn interaction. 
On the other hand, they  also draw attention on the fact that  for $^{135}$Sb  the latter interaction
does better than the KH208 interaction (see Ref. \cite{sher05a} for details).  On these grounds,
their conclusion \cite{sher05} is 
that a consistent Hamiltonian for the three nuclei $^{134}$Sn,  $^{134}$Sb, and  $^{135}$Sb has yet to be found.

Our point of view is quite different.
In fact, we have shown here that our realistic effective interaction gives a good description of $^{134}$Sb  while   
the results presented in Refs. \cite{cor05,cove} evidence that  it is also able to reproduce the spectroscopic properties 
of $^{135}$Sb and $^{134}$Sn. In particular, in \cite{cor05} we have reported the diagonal matrix elements of the
proton-neutron effective interaction for the 
$\pi g_{7/2} \nu f_{7/2}$ and   $\pi d_{5/2} \nu f_{7/2}$ configurations  and pointed out
their  crucial role for  the structure of the low-lying $5/2^{+}$ state in $^{135}$Sb.
As for $^{134}$Sn, it is worth noting that the results obtained in \cite{cove}
are not significantly different 
from those of \cite{cor02,cove?}, where slightly different effective interactions were employed.   

In this context, it is interesting to try to understand what 
makes our proton-neutron matrix elements   appropriate to the description of  the
multiplets in $^{134}$Sb, in particular  the very small energy spacing between  
the $0^{-}$ and the $1^{-}$ states. To this end, we have performed an analysis of the various contributions to
 our effective interaction, focusing
attention on  the $\pi g_{7/2} \nu f_{7/2}$ configuration. In fact,  in this case a direct relation can be established 
between the diagonal matrix elements and the behavior of the multiplet, since its members are, as mentioned above, 
almost pure.
We start by reporting in Fig. 2 the behavior of the diagonal matrix elements of the effective 
interaction as a a function of $J$.  
We see, as expected, that  the pattern is quite similar to that of the corresponding multiplet in Fig. 1 and  
the matrix elements for the $0^{-}$ and the $1^{-}$ states are almost equal.

\begin{figure}[H]
\begin{center}
\includegraphics[scale=0.55,angle=0]{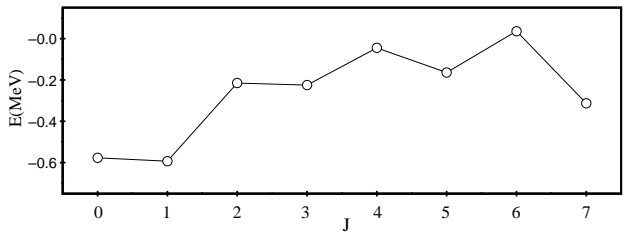}
\end{center}
\caption{Diagonal matrix elements of the two-body effective 
interaction for the $\pi g_{7/2} \nu f_{7/2}$ configuration.}
\end{figure}

As mentioned above, our effective interaction  is calculated within the framework of 
a $\hat{Q}$-box folded-diagram method. In particular, the $\hat{Q}$ box
is composed of first- and second-order diagrams in the $V_{\rm low-k}$ derived from
the CD-Bonn potential. In other words, the matrix elements of the effective interaction contain
the $V_{\rm low-k}$ plus additional  terms which take into account
core-polarization effects arising from $1p-1h$ (``bubble'' diagram) and $2p-2h$ excitations.
They also include the so-called ladder diagrams, which must compensate for the excluded configurations
above the chosen model space.
We may point out  that the effective interaction is obtained by summing the $\hat{Q}$-box folded-diagram series
\cite{shur}.

In Fig. 3 we show the $\pi g_{7/2} \nu f_{7/2}$ 
matrix elements of the $V_{\rm low-k}$ as a function of $J$ together with  the
second-order two-body contributions.
From the inspection of this figure we see that the incorrect behavior
of the $V_{\rm low-k}$ 
matrix elements is ``healed'' by
the $V_{1p1h}$, $V_{2p2h}$  and $V_{\rm ladder}$ corrections.
In particular it appears that a crucial role is played by the
bubble diagram, especially as regards the
position of the $1^-$ state. In this connection, we may mention that the effect of
the bubble renormalization on the $0^{-}-1^{-}$ splitting was also pointed out in Ref.~\cite{sher05}
with regard to the Kuo-Herling interaction. It is worth noting that in our calculations the folding procedure
turns out to provide
a common attenuation of all matrix elements, which does not affect
the overall behavior of the multiplet.  

\begin{figure}[H]
\begin{center}
\includegraphics[scale=0.55,angle=0]{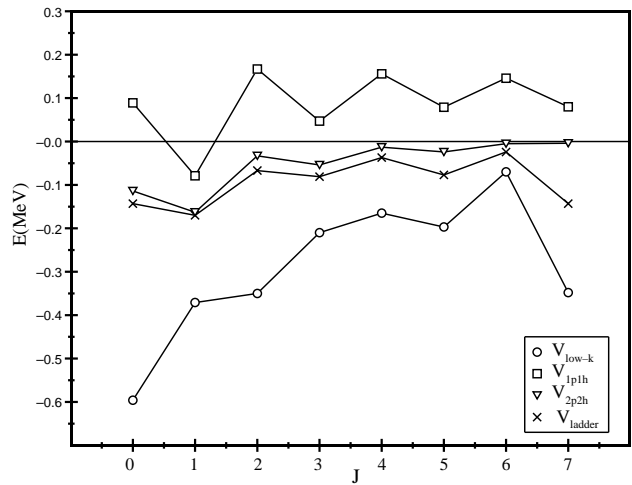}
\end{center}
\caption{Diagonal matrix elements of $V_{\rm low-k}$ and contributions 
from the two-body second-order diagrams for the $\pi g_{7/2} \nu
f_{7/2}$ configuration. See text for comments.}
\end{figure}

To summarize, we have shown that our realistic shell-model calculation for $^{134}$Sb leads to a
very good agreement with experiment for the 
$\pi g_{7/2} \nu f_{7/2}$ multiplet  as well as for the two observed   members of 
the $\pi d_{5/2} \nu f_{7/2}$ multiplet.  
This outcome, when considered along with the results we have obtained for $^{134}$Sn and $^{135}$Sb, evidences 
the merit of our effective interaction in describing the properties of $^{132}$Sn neighbors 
with neutrons beyond the 82 shell. 
We have also examined the various terms of our 
effective interaction, in order to understand their relative importance in the resulting final matrix elements.
We have evidenced the importance of the renormalizations one has to introduce to
account for the limited size of the chosen model space, in particular those arising from the $1p-1h$ excitations.  
Based on these results, we may conclude that to
explain the presently available data on  neutron-rich nuclei beyond $^{132}$Sn there is no
need to invoke shell-structure modifications.

\begin{acknowledgments}
This work was supported in part by the Italian Ministero
dell'Istruzione, dell'Universit\`a e della Ricerca  (MIUR).
\end{acknowledgments}


\begin{thebibliography}{9}
\bibitem{nord} L. W. Nordheim, Phys Rev. {\bf 78}, 294 (1950); Rev. Mod Phys. {\bf 23}, 322 (1951).
\bibitem{sher02} J. Shergur {\it et al.}, Phys. Rev. C {\bf 65},
034313 (2002).
\bibitem{sher05a} J. Shergur {\it et al.}, Phys. Rev. C {\bf 72},
024305 (2005).
\bibitem{sher05} J. Shergur {\it et al.}, Phys. Rev. C {\bf 71},
064321 (2005).
\bibitem{chou92} W. T. Chou and E. K. Warburton, Phys. Rev. C {\bf 45}, 1720 (1992).
\bibitem{sark01} S. Sarkar and M. S. Sarkar, Phys. Rev. C {\bf 64}, 014312 (12001).
\bibitem{korg02} A. Korgul {\it et al.}, Eur. Phys. J. A {\bf 15}, 181 (2002).
\bibitem{warb91} E. K. Warburton and B. A. Brown, Phys. Rev. C {\bf 43}, 602 (1991).
\bibitem{kuo71} T. T. S. Kuo and G. H. Herling, US Naval Research Laboratory Report No. 2258, 1971 (unpublished).
\bibitem{cor05} L. Coraggio, A. Covello, A. Gargano, and N. Itaco, Phys. Rev.
C, {\bf 72} 057302 (2005).
\bibitem{cove} A. Covello, L. Coraggio, A. Gargano, and N. Itaco,
BgNS Transactions, in press. 
\bibitem{ei} T. T. S. Kuo and E. Osnes, {\it Lecture Notes in Physics}, Vol. 
364, (Springer-Verlag, Berlin, 1990).
\bibitem{mac01} R. Machleidt, Phys. Rev. C {\bf 63}, 024001 (2001).
\bibitem{bog02} S. Bogner, T. T. S. Kuo, L. Coraggio, A. Covello, and N. Itaco, Phys. Rev.
C {\bf 65}, 051301 (2002).
\bibitem{and97} F. Andreozzi, L. Coraggio, A. Covello, A. Gargano, T. T. S. Kuo, and A. Porrino, 
Phys. Rev. C {\bf 56}, R16 (1997).
\bibitem{cor02} L. Coraggio, A. Covello, A. Gargano, and N. Itaco, Phys. Rev.
C {\bf 65}, 051306(R) (2002).
\bibitem{oxba}B. A. Brown, A. Etchegoyen, and W. D. M. Rae, The computer code OXBAH, MSU-NSCL, Report No. 534.
\bibitem{audi03} G. Audi, A. H. Wapstra, and C. Thibault, Nucl. Phys. A {\bf 729}, 337 (2003).
\bibitem{foge99} B. Fogelberg, K. A. Mezilev, H. Mach, V. I. Isakov, and J. Slivova,
Phys. Rev. Lett. {\bf 82}, 1823 (1999).
\bibitem{nndc} Data extracted using the NNDC On-line Data Service from the XUNDL database, file 
revised as of November 8, 2005.
\bibitem{garga04} A.  Gargano, Eur. Phys. J. A {\bf 20}, 103 (2004).
\bibitem{brown05} B. A. Brown, N. J. Stone, J. R. Stone, L. S. Towner, and M. Hjorth-Jensen, 
Phys Rev, C {\bf 71}, 044317 (2005) and {\bf 72}, 029901(E) (2005).
\bibitem{cove?} A. Covello, L. Coraggio, A. Gargano, and N. Itaco,
Journal of Physics: Conference Series {\bf 20}, 137 (2005).  
\bibitem{shur} J. Shurpin, T. T. S. Kuo, D. Stottman, Nucl. Phys. A  {\bf 408}, 310 (1983).

\end{thebibliography}
\end{document}